\documentclass[aps,prb,twocolumn,showpacs,preprintnumbers]{revtex4}

\usepackage {epsfig,color}
\usepackage{bm}
\usepackage {graphicx}
\usepackage[T2B]{fontenc}

\begin{document}

\title {Density Functional Theory and Generalized Tight-Binding combined method for Hubbard fermion-phonon coupling study in strongly correlated LSCO-system}

\author{J. Spitaler$^{1,2}$}
\author{E.I. Shneyder$^3$}
\author{E.E. Kokorina$^4$}
\author{I.A. Nekrasov$^4$}
\email {shneyder@iph.krasn.ru}
\author{V.A. Gavrichkov$^{3}$}
\author{C. Ambrosch-Draxl$^1$}
\author{S.G. Ovchinnikov$^{3}$}

\affiliation{$^1$Chair of Atomistic Modelling and Design of Materials, Montanuniversit\"{a}t Leoben,
Franz-Josef-Stra\ss e~18, A-8700 Leoben, Austria}
\affiliation{$^2$Material Center Leoben, Rosegger-Stra\ss e~18, A-8700 Leoben, Austria}
\affiliation{$^3$Kirensky Institute of Physics SB RAS, 660036 Krasnoyarsk, Russia}
\affiliation{$^4$Institute of Electrophysics UB RAS, Amundsena Str. 106, 620016 Yekaterinburg, Russia}
\date{\today}


\begin {abstract}
We present {\it ab initio} results for the electron-phonon interaction of the $\Gamma$-point phonons in the tetragonal high-temperature phase of La$_2$CuO$_4$. Eigenfrequencies and eigenvectors for the symmetry-allowed phonon modes are calculated with the full-potential
augmented plane wave+local orbitals method using the frozen phonon approach. It is found that the $\Gamma$-point phonons with the strongest electron-phonon interaction are the A$_{2u}$ modes with 236 cm$^{-1}$, 131 cm$^{-1}$ and 476 cm$^{-1}$. To take effect of strong electron on-site interaction into account we use generalized tight-binding method that results in the interaction of phonons with Hubbard fermions forming quasiparticle's band structure. Finally, the matrix elements of Hubbard fermion-phonon interaction and their reduction due to strong electron correlation are obtained.

\end {abstract}


\maketitle

\section{Introduction}\label{intro}
All physical properties of solids are affected by electron-phonon interaction (EPI). The most powerful manifestation of EPI is superconductivity in metals. In high-$T_c$ compounds effect of EPI on superconducting pairing  is also well pronounced~\cite{Maksimov10,Kulic05,Maksimov00} however its crucial role is under debate. The interpretation of underlying superconducting mechanisms in these materials is complicated since intrinsic strong electron correlations can both induce interactions which compete with EPI in pairing and modify electron-phonon coupling along with properties caused by it. To disentangle electron and phonon contributions in correlated materials new experimental and theoretical tools are wanted.

Recently an attempt to separate the electronic and phononic glue in high-T$_c$ superconductor has been carried out using femtosecond spectroscopy on Bi$_2$Sr$_2$Ca$_{0.92}$Y$_{0.08}$Cu$_2$O$_{8+\delta}$ crystals with simultaneous time and frequency resolution\cite{DalConte12}. Analyzing the different temporal evolution of the electronic and phononic contributions to the total pairing interaction $\Pi \left( \Omega  \right)$ the authors claim the dominant role of the electronic mechanisms of pairing $\left( { \sim 80\% } \right)$ with minor $\left( { \sim 20\% } \right)$ contribution of the EPI. This very interesting approach nevertheless requires more deep theoretical analysis. In the Eliashberg theory\cite{Eliashberg60} the total interaction determines both the pairing coupling and the renormalization of the electron dispersion in the normal phase. Our meanfield analysis\cite{Shneyder10} of the gap equation and isotope effect with simultaneous accounting of strong electron correlation and EPI results in the approximately equal contributions of the magnetic and EPI coupling in high-T$_c$ cuprates. However developed for extended Hubbard model theory similar to the Migdal-Eliashberg one with self-energy defined in non-crossing approach reveals only small contributions of phonons to superconducting pairing beside spin fluctuations induced by strong kinematic interaction~\cite{Plakida2013}. Thus the problem of disentangling the electronic and phononic glue is far from being completed.

Theoretical study of the systems in which electron-electron interaction and electron-phonon one are nonnegligible demands to define the band structure, electron self-energy and polarization operators self-consistently. It is obviously that competition of these interactions can result~\cite{Plakida2013,Sadovskii11,Capone2010} in different physics depending on both considered model and relevant physical parameters therefore realistic approaches are required. Since {\it ab initio} density functional theory (DFT)\cite{Hohenberg,KohnSham} augmented with the local density approximation (LDA)\cite{Perdew92} or generalized gradient approximation fails to describe strongly correlated systems new methods are developed.\cite{Savrasov03,Kotliar2013} Linear response approach based on dynamical mean field theory (DMFT)\cite{DMFT96} and LDA enables\cite{Savrasov03} to study the lattice dynamics and structural stability of correlated materials if correlation effects are mainly local and self-energy is dominated by the frequency dependence. Method combining the LDA linear response calculations and a few supercell calculations based on GW approximation~\cite{GW1998} and screened hybrid functional DFT provides description of electronic structure and lattice dynamical properties of materials if self-energy has mostly the momentum dependence.

However these methods are often  highly computationally demanding. For normal metals advantages of realistic DFT and model consideration have been successfully combined in hybrid scheme\cite{Maksimov97}. The later utilizes {\it ab initio} calculated\cite{Savrasov94,Savrasov96,2Savrasov96} matrix elements of EPI and spectral functions which characterize electron-phonon scattering in the equations of developed\cite{Scalapino69,Rainer86,Allen82,Dolgov82} many-body theory explaining properties of solids caused by EPI in normal and superconducting states. To justify the hybrid methods for materials with strong electron correlations we propose to convert characteristics of EPI obtained in single-electron LDA picture in parameters of generalized tight-binding method (GTB)~\cite{Ovchinnikov89} which describe interaction of phonons with Hubbard fermions forming quasiparticle's band structure. The GTB approach has been proposed earlier to study the electronic structure of strong correlated electron systems as a generalization of Hubbard ideas for the realistic multiband Hubbard-like models. This method combines the exact diagonalization of the intracell part of the Hamiltonian, construction of the Hubbard operators on the basis of the exact intracell multielectron eigenstates, and the perturbation treatment of the intercell hoppings and interactions. A similar approach to the 3-band $p-d$ model of cuprates\cite{Emery87,Varma87} is known as the cell perturbation theory\cite{Lovtsov91,Jefferson92,Schuttler92,Plakida99}.

The rest of the paper is organized in the following way: section~\ref{computation} reports the computational details for the augmented plane wave+local orbitals method, section~\ref{phonons} contains the description of phonon eigenvectors and eigenfrequencies for La$_2$CuO$_4$ obtained from {\it ab-initio} calculations. These data are defined more precisely towards previous calculations. The resulting EPI matrix elements are presented in section~\ref{EPI}, the scheme of proposed combined method for Hubbard fermion-phonon coupling study  and a discussion of the obtained results are given in section~\ref{EPI+GTB}.

\section{Computational details}\label{computation}

In order to obtain an optimized lattice constants and atomic positions as a starting point for
frozen-phonon calculations, the augmented plane wave + local orbital (FP-APW+lo)
method\cite{Sjoestedt00} implemented in the WIEN2k code\cite{wien2k} has been used.
This method provides the most accurate way to treat crystals within density
functional theory. At this stage exchange and correlation effects have been treated within the local density approximation. The atomic sphere radii
$R_{\rm MT}$ have been chosen as 2.4~a.u.~for La, 1.875~a.u.~for Cu, and
1.575~a.u.~for the O atoms. For the wavefunctions, a plane wave cutoff $K_{\rm
max} = 4.44$ has been used, which corresponds to about 970 basis  functions and
$R_{\rm MT}\times K_{\rm max}$ values of 10.67 for La, 8.33 for Cu and 7.0 for
O, respectively. We used 512 ${\bf k}$ points in the full Brillouin zone (BZ)
for the self-consistency  cycles,  which yields 56 points in the irreducible
wedge. The plane wave cutoff for expanding the charge density and potential in
the  interstitial region, $G_{\rm max}$, was 14.

All degrees of freedom, {\emph i.e.}, the atomic positions and the
lattice constants, have been optimized starting from the experimental ones
reported in Ref.~\cite{Longo73}. The space group of La$_2$CuO$_4$ in the
tetragonal phase is $I 4/m m m$ (no.~139).
The structure with the lowest total energy is found for a=3.72~\AA, c=12.98~\AA,
{\emph i.e.}, the volume is decreased by about 5\%.
The positions of La atom and apex oxygen O2 are obtained as (0, 0, 0.3616 c) and (0, 0, 0.184 c),
respectively, which implies that the $z$ coordinate of the apex oxygen changes by 0.03~\AA,
while the position of La is not affected by the optimization. LDA band dispersions calculated for given crystal structure by means of FP+APW+lo are presented in fig.1.

\section{$\Gamma$ point phonons}\label{phonons}

\begin{table*}[htb]
\begin{center}
\begin{tabular}{c@{\hspace{3mm}}c@{\hspace{5mm}}*{12}{r@{\hspace{3mm}}}}
\hline
\hline
 Mode & $\omega$ & La$_x$ & La$_y$ &La$_z$  & Cu$_{x,y}$ &  Cu$_z$ & O1$_x$ & O1$_y$ & O1$_z$ & O2$_x$ & O2$_y$& O2$_z$ \\
      & (cm$^{-1}$)    &        &        &        &        &        &        &        &        &        &        &        \\
\hline

A$_{1g}$  &   415    &   --   &   --   &  0.08  &   --   &    --   &   --   &   --   &   --   &   --   &   --   &  1.00 \\
      &   232    &   --   &   --   &  1.00  &   --   &    --   &   --   &   --   &   --   &   --   &   --   & -0.08 \\[2mm]

A$_{2u}$ &   476    &   --   &   --   & -0.10  &   --   &   0.06  &   --   &   --   & -0.57  &   --   &   --   &  0.81 \\
      &   236    &   --   &   --   &  0.38  &   --   &  0.10  &   --   &   --   & -0.77  &   --   &   --   & -0.50 \\
      &   131    &   --   &   --   &  0.44  &   --   & -0.89  &   --   &   --   &  0.02  &   --   &   --   &  0.14 \\[2mm]

B$_{2u}$ &   227    &   --   &   --   &   --   &   --   &   --   &   --   &   --   &  1.00  &   --   &   --   &   -- \\[2mm]

E$_g$   &   209    & -0.28  &  0.28  &   --   &   --   &   --   &   --   &   --   &   --   &  0.96  & -0.96  &   -- \\
      &    73    &  0.96  & -0.96  &   --   &   --   &   --   &   --   &   --   &   --   &  0.28  & -0.28  &   -- \\[2mm]

E$_u$   &   727    &  0.00  &  0.00  &   --   & 0.25  &   --   &  0.09  & -0.93  &   --   &  0.02  &  0.02  &   -- \\
      &   341    &  0.05  &  0.05  &   --   &  0.27  &   --   & -0.92  &  0.05  &   --   & -0.04  & -0.04  &   -- \\
      &   210    &  0.19  &  0.19  &   --   & -0.59  &   --   & -0.34  & -0.34  &   --   &  0.05  &  0.05  &   -- \\
      &    80    &  0.14  &  0.14  &   --   & -0.02  &   --   &  0.07  & -0.04  &   --   & -0.69  & -0.69  &   -- \\[2mm]
\hline
\hline
\end{tabular}
\caption{Calculated eigenfrequencies (in cm$^{-1}$) and eigenvectors of the $\Gamma$
point modes of La$_2$CuO$_4$.}
\label{eigenmodes}
\end{center}
\end{table*}

\begin{table*}[htb]
\begin{center}
\begin{tabular}{c@{\hspace{3mm}}c@{\hspace{8mm}}*{6}{c@{\hspace{3mm}}}c}

\hline
\hline
      & This & \multicolumn{6}{c}{Literature}\\
      & work & LDA\cite{Wang99}
                      & LDA\cite{Singh96}
                                & LDA\cite{Cohen89}
                                          & IR                                    & Neutron\cite{Pintschovius91}  & Raman\cite{Burns90} \\
\hline
 A$_{1g}$ &  415 & 375    &   390   &  415    & --                                    &     427     & 433      \\
      &  232 & 202    &   215   &  224    & --                                    &     227     & 226      \\[2mm]

A$_{2u}$ &  476 & 441    &   446   &   --    & 500\cite{Henn97}, 501\cite{Collins89} &     497     &   --     \\
      &  236 & 182    &   197   &   --    & 235\cite{Henn97}, 342\cite{Collins89} &     251     &   --     \\
      &  131 & 132    &   119   &   --    & 135\cite{Henn97}, 242\cite{Collins89} &     149     &   --     \\[2mm]

B$_{2u}$ &  227 & 193    &   201   &  293    &  --                                   &     270     &   --     \\[2mm]

E$_g$   &  209 & 201    &   212   &  233    &  --                                   &     241     &   --     \\
      &   73 &  26    &    15   &         &  --                                   &      91     &   --     \\[2mm]

E$_u$   &  727 & 630    &   650   &    --   &   695\cite{Collins89}                 &     684     &   --     \\
      &  341 & 319    &   312   &    --   &   360\cite{Collins89}                 &     354     &   --     \\
      &  209 & 147    &   146   &    --   &   140\cite{Collins89}                 &     173     &   --     \\
      &   80 &  22    &   75i   &   39    &                                       &     126     &   --     \\[2mm]
\hline
\hline

\end{tabular}
\caption{Calculated eigenfrequencies (in cm$^{-1}$) of the $\Gamma$ point phonons compared to literature.
Papers~\cite{Wang99,Singh96,Cohen89} refer to DFT calculations using LDA,
papers~\cite{Henn97} and \cite{Collins89} to infrared measurements, and
paper~\cite{Burns90} to a Raman experiment. The frequencies reported for neutron diffraction are extracted~\cite{Pintschovius91} from data of La$_{1.9}$Sr$_{0.1}$CuO$_{4}$ at 295 K.
 }
\label{phonon_comparison}
\end{center}
\end{table*}

Tab.~\ref{eigenmodes} shows the eigenfrequencies and eigenvectors for all
$\Gamma$ point phonons of La$_2$CuO$_4$, while Tab.~\ref{phonon_comparison}
compares the eigenfrequencies with previous ab-inito calculations
~\cite{Wang99,Singh96,Cohen89} and results from infrared (IR) measurements,
~\cite{Henn97,Collins89} neutron diffraction\cite{Pintschovius91} and Raman
scattering experiments.~\cite{Burns90} The phonon modes have been obtained in the
frozen phonon approximation. Details about the procedure are found, for example,
in Ref.~\cite{Ambrosch02}. For each degree of freedom 4 displacements, two
in  positive and two in negative direction, have been calculated. A linear fit
of the resulting forces as a function of displacement has been used to set up
the dynamical matrix, which yields the phonon eigenfrequencies and eigenvectors.

In all three theoretical works~\cite{Wang99,Singh96,Cohen89} the LAPW method together with the LDA for the exchange-correlation potential was used for the groundstate computations.
Compared to these calculations we have used a considerably larger set of basis
functions (about 950 compared to 750 in Refs.~\onlinecite{Wang99,Singh96} and
650 in Ref.~\onlinecite{Cohen89}), and the generally more accurate APW+l.o.~scheme.
Regarding the lattice constants, Refs.~\onlinecite{Wang99} and
\onlinecite{Singh96} have used the experimental ones, while
Ref.~\onlinecite{Cohen89} has optimized both the volume and the c/a ratio
yielding the same geometry as ours. For the phonon calculations,
Refs.~\onlinecite{Singh96} and  Refs.~\onlinecite{Cohen89} applied the frozen phonon method, while  the results in Ref.~\onlinecite{Wang99} are based on a
linear-response calculations.

Our frequencies for the A$_{1g}$~phonons are---as the ones presented by Cohen et
al.\cite{Cohen89}---in
excellent agreement with experiment. A comparison of these results with the ones obtained without geometry optimization\cite{Wang99,Singh96} demonstrates that the optimized  geometry
improves the frequencies considerably. A similar situation is found for the
A$_{2u}$~modes, where the agreement with experiment is again improved significantly
and gives very good results. For the B$_{2u}$~mode the situation is different: our
frequency of 227~cm$^{-1}$ is about 10\% higher than the one obtained with
experimental geometry\cite{Wang99,Singh96} and thus closer to the result of
270~cm$^{-1}$ obtained by neutron diffraction,\cite{Pintschovius91} while Cohen et
al.\cite{Cohen89} report 293~cm$^{-1}$ for the same mode. This is rather surprising,
since the latter used a very similar method and the same geometry as it has been
used in our calculations.
Analyzing, finally, the results for the E$_g$ and E$_u$ modes,
it turns out that our calculations have especially improved the results
for the low-frequency modes of either species, where former calculations yielded
either extremely low or even imaginary frequencies. This indicates that for the
low-energy features the higher accuracy resulting  from the larger basis set is
especially important.

The eigenvectors are similar to the ones presented in Ref.~\onlinecite{Wang99}.
A major difference is found for the A$_2u$~mode with 236~cm$^{-1}$, where the
eigenvector reported by Ref.~\onlinecite{Wang99} has a smaller contribution of
La, but a much larger component of the Cu atom. The frequency presented in
this reference is only 182~cm$^{-1}$, which indicates that the larger mass of La
compared to Cu,  which should lead to a lower frequency, is more than
compensated by the effect of geometry optimization, where latter yields
a result much closer to experiment.

\section{Electron phonon interaction}\label{EPI}

The electronic band structure for the equilibrium and changed ionic positions of all
$\Gamma$ point phonons of La$_2$CuO$_4$ has been calculated. Then in each case the difference between the Kohn-Sham eigenvalues of the distorted and undistorted system has been extracted. In order to compare the electron-phonon coupling parameters for the different phonon modes with each other, the displacements are normalized to obtain the dimensionless phonon coordinate $Q$ defined as\cite{Spitaler07}
\begin{equation}
Q\sqrt {\frac{\hbar }{{{M_\alpha }{\omega _\beta }}}} {{\bf{e}}_{\alpha \beta }} = {\bf{u}}_\beta ^\alpha.
\end{equation}
Here, $\omega _\beta$ is the eigenfrequency of the considered mode, $M_\alpha$ is the mass of ion $\alpha$ and ${\bf{u}}_\beta ^\alpha$ is the corresponding real displacement.

\begin{table*}[htb]
\begin{center}
\begin{tabular}{cc@{\hspace{3mm}}*{6}{r@{\hspace{3mm}}}}
\hline
\hline
          &         &    \multicolumn{6}{c}{Band} \\
          &       & Cu-d$_{x^2-y^2}$ & Cu-d$_{3z^2-r^2}$ & O1-2p$_x$ &O1-2p$_y$ & O2-p$_z$(1)&O2-p$_z$(2)\\

\hline
 A$_{1g}$ & 415   &  0.137   &  0.137   &   0.776 &   0.776 &   2.147 &  1.462 \\
          & 232   &  0.264   &  0.616   &   0.459 &   0.459 &   1.398 &  0.713 \\[2mm]

A$_{2u}$  & 476   &  0.517   &  0.765   &   1.335 &   1.335 &  2.755 &   1.042 \\
          & 236   &  2.701   &  3.913   &   8.806 &   8.806 &  3.791 &   1.054 \\
          & 131   &  0.829   &  1.557   &   3.132 &   3.132 &  3.332 &   0.692 \\[2mm]

B$_{2u}$  & 227   &  0.082   &  0.519   &   0.585 &   0.585 &  1.029 &   0.239 \\[2mm]

E$_g$     & 209   &  0.036   &  0.125   &   0.073 &   0.073 &  0.048 &   0.273 \\
          &  73   &  0.015   &  0.059   &   0.278 &   0.254 &  0.039 &   0.161 \\[2mm]

E${_u}$   & 727   &  0.320   &  0.232   &   1.038 &  1.093 &   0.784 &   0.342 \\
          & 341   &  0.198   &  0.568   &   1.366 &  1.568 &   1.213 &   0.568 \\
          & 209   &  0.053   &  0.059   &   0.133 &  0.149 &   0.156 &   0.108 \\
          &  80   &  0.060   &  0.213   &   0.132 &  0.118 &   0.155 &   0.198 \\
\hline
\hline
\end{tabular}
\caption{Absolute value (in eV) of electron-phonon interaction parameters at ${\bf k}=0$ point of Brillouin zone for the different bands and $\Gamma$ point modes.}
\label{epimat}
\end{center}
\end{table*}
The results for ${\bf k}=0$ point of Brillouin zone are given in Table~\ref{epimat}. We consider ionic position dependence of the 6 bands with predominant contribution of Cu-$d_{x^2-y^2}$ and Cu-$d_{3z^2-r^2}$ bands, $p_x$ and $p_y$ bands of the in-plane oxygen O1, and $p_z$ bands of the apex oxygen O2. Such set of bands corresponds to 5-band $p-d$ model and provide proper description of LDA bands near Fermi level. The electron-phonon interaction is very different for the different phonon modes: it is small for all bands for both E$_g$ modes as well as for E$_u$ modes with 210 cm$^{-1}$ and 80 cm$^{-1}$ and already an order of magnitude larger for the O1-$p_x,p_y$ bands in case of the E$_u$ modes with 727 and 341~cm$^{-1}$. All of the A$_{1g}$  modes exhibit considerable coupling to the O2-$p_z$ bands and O1-$p_x,p_y$ bands. The same four bands and also Cu-$d_{3z^2-r^2}$ band are the most affected ones in case of the B$_{2u}$ phonon modes. The strongest EPI show three A$_{2u}$ modes which especially alter the O1-$p_x,p_y$, O2-$p_z$, and Cu-$d_{3z^2-r^2}$ levels. The origin of this strong coupling is a poor screening of the Coulomb potential perpendicular to the conducting CuO$_2$ layers\cite{Falter97} that results in a strong modulation of the Madelung potential by $c$-axes phonons. A direct proof of strong coupling between electrons and ionic displacements along the $c$-axes was the colossal heat expansion of La$_2$CuO$_4$ under high-power femtosecond light irradiation.\cite{Gedik07} In this experiment a sudden increase of the $c$-lattice parameter induced by photodoped holes has been observed. The  atomic displacements of the modes with the largest EPI, {\emph i.e.}, the A$_{2u}$ modes with 236~cm$^{-1}$, 131~cm$^{-1}$, and 476~cm$^{-1}$ are presented in Figs.~2.

\section{Hubbard fermion-phonon coupling study in strongly correlated LSCO-system}\label{EPI+GTB}

In conventional metals electron-phonon interaction is of the form
\begin{equation}
\label{EPI_Hamiltonian}
{H_{EPI}} = \sum\limits_{{\bf{k}},{\bf{q}},\nu ,\lambda,\sigma } {{g_\rho }\left( {{\bf{k}},{\bf{q}}{\rm{;}}\nu } \right)c_{{\bf{k}} - {\bf{q}},\lambda,\sigma }^ \dag {c_{{\bf{k}},\lambda,\sigma }}\varphi _{\bf{q}}^\nu }
\end{equation}
where $\varphi _{\bf{q}}^\nu  = \left( {{b_{{\bf{q}},\nu }} + b_{ - {\bf{q}},\nu }^ \dag } \right)$ and ${b_{{\bf{q}},\nu }}\left( {b_{ - {\bf{q}},\nu }^ \dag } \right)$ is destruction (creation) operator of phonon of branch $\nu$ and momentum ${\bf q}$; operators ${c_{{\bf{k}},\lambda,\sigma }}$ and $ {c_{{\bf{k}},\lambda,\sigma }^ \dag } $ describe destruction and creation of electron with spin $\sigma$ and initial momentum $\bf{k}$ in band $\lambda$; ${{g_\lambda }\left( {{\bf{k}},{\bf{q}}{\rm{;}}\nu } \right)}$ is matrix element of interaction between phonon and band electron. To properly evaluate the matrix element of EPI in correlated material we should take into consideration that band structure of such system is not given by single electron picture of DFT augmented with LDA or GGA method. Indeed, band structure of strongly correlated electrons in cuprates looks like quasiparticle bandstructure of Hubbard fermions results from multi electron approach like LDA+DMFT\cite{Anisimov97,Lichtenstein98,Held2001,Kotliar2006} or LDA+GTB\cite{Korshunov2012}. For example, LDA results in incorrect metallic state of La$_2$CuO$_4$ while GTB method reproduces charge transfer insulator for undoped system and strong spectral weight redistribution between Hubbard subbands with hole doping in underdoped La$_{2-x}$Sr$_x$CuO$_4$~\cite{Gav2001}.

From the very beginning the GTB method has been suggested for the Mott-Hubbard insulators like transition metal oxides to extend the microscopic band structure calculations and take the strong electron correlations into account. As any other cluster perturbation theory, the GTB method starts with the exact diagonalization of the intracell part of the multielectron Hamiltonian and treats the intercell part by a perturbation theory. The exact diagonalization of the intracell Hamiltonian results in complete set of orthogonal and normalized eigenstates $\left\{ {\left| p \right\rangle } \right\}$ and allow us to construct the Hubbard operators $X_f^{pq} = \left| p \right\rangle \left\langle q \right|$. Due to definition of Hubbard operators any local operators can be presented as a linear combination of $X$-operators, for example electron destruction operator in the cell $f$ with the band index $\lambda$ takes a form
\begin{eqnarray}
\label{def_Hubb_op}
{c_{f,\lambda ,\sigma }}& =& \sum\limits_{p,p'} {\left| p \right\rangle \left\langle p \right|} {c_{f,\lambda ,\sigma }}\left| p' \right\rangle \left\langle p' \right| = \nonumber \\
 &=&\sum\limits_{p,p'} {{\gamma _{\lambda ,\sigma }}\left( {pp'} \right)X_f^{pp'}}.
\end{eqnarray}
Equation (\ref{def_Hubb_op}) shows clearly the difference in the Fermi type quasiparticle description in the single electron language and in the multielectron one. The operator ${c_{f,\lambda ,\sigma }}$ decreases the number of electrons by one for all sectors of the Hilbert space simultaneously, while the ${X_f^{pq}}$ operator describe the partial process of electron removing in the $\left( {N + 1} \right)$-electron configuration $\left| p \right\rangle $ with the final N-electron configuration $\left| p' \right\rangle $. The matrix element $\left\langle p \right|{c_{f,\lambda ,\sigma }}\left| {p'} \right\rangle $ gives the probability of such process. The splitting of electron stated by equation (\ref{def_Hubb_op}) on different Hubbard fermions and following spectral weight redistribution over these quasiparticles are the underlying effects of band structure formation in correlated systems. It is obviously that applying the presentation of electron (\ref{def_Hubb_op}) to Hamiltonian (\ref{EPI_Hamiltonian}) will result in the interaction of phonons with Hubbard fermions forming quasiparticle's band structure.

To consider the problem in details we proceed with microscopical model of La$_2$CuO$_4$ system which reflects its  chemical structure and contains proper description of its low energy physics~\cite{Gav2000}:
\begin {eqnarray}
\label{pd_Hamiltonian}
H_{pd} &=&\sum\limits_ {f, \lambda, \sigma}(\epsilon_ {\lambda}-\mu) n_ {f,
\lambda, \sigma} + \sum\limits_ {f \neq g} \sum\limits_ {\lambda, \lambda',
\sigma} T_ {f g}^ {\lambda \lambda'} c_ {f, \lambda, \sigma}^\dag c_ {g,
\lambda', \sigma}+ \nonumber \\ &+& \frac {1}{2}\sum\limits_ {f, g, \lambda,
\lambda'} \sum\limits_ {\sigma_ {1,2,3,4}} V_ {f g}^{\lambda \lambda'} c_{f, \lambda, \sigma_1}^\dag c_{f, \lambda, \sigma_3} c_{g, \lambda', \sigma_ 2}^\dag c_{g, \lambda', \sigma_4}.
\end {eqnarray}
Here $c_{f, \lambda, \sigma}$ is the annihilation operator in Wannier representation of the  hole at orbital $\lambda$ with spin $\sigma$,  and $n_{f, \lambda, \sigma}=c_{f, \lambda,
\sigma}^\dag c_{f, \lambda, \sigma}$. Index $f$ enumerates copper and oxygen sites, index $\lambda$ runs through $d_{x^2-y^2}$ and $d_{3z^2-r^2}$ orbitals on copper,
$p_x$ and $p_y$ atomic orbitals on the plane-oxygen sites and $p_z$ orbitals
on the apical oxygen; $\epsilon_{\lambda}$ - single-electron energy of the
atomic orbital $\lambda$.  $T_{f g}^{\lambda \lambda'}$ includes matrix
elements of hoppings between copper and  oxygen ($t_{pd}$ for hopping $d_{x^2-y^2}
\leftrightarrow p_x,p_y$; $t_{pd}/\sqrt{3}$ for  $d_{3z^2-r^2} \leftrightarrow
p_x,p_y$; $t'_{pd}$ for $d_{3z^2-r^2} \leftrightarrow p_z$) and between  oxygen and
oxygen ($t_{pp}$ for hopping $p_x \leftrightarrow p_y$; $t'_{pp}$ for
hopping $p_x,p_y \leftrightarrow p_z$). The Coulomb matrix elements  $V_{f
g}^{\lambda \lambda'}$ includes intraatomic Hubbard repulsions of two
holes  with opposite spins on one copper and oxygen orbital ($U_d$, $U_p$),
between different  orbitals of copper and oxygen ($V_d$, $V_p$), Hund
exchange on copper and oxygen  ($J_d$, $J_p$) and the nearest-neighbor
copper-oxygen Coulomb repulsion $V_{pd}$. The {\it ab initio} hopping parameters and single electron energies of the Hamiltonian (\ref{pd_Hamiltonian}) have been obtained~\cite{LDA+GTB} in Wannier Function projection procedure~\cite{wf_Anisimov}, Coulomb parameters have been defined~\cite{Sorella} in constrained LDA supercell calculations~\cite{constrainedLDA}.

The procedure of exact diagonalization of unit cell Hamiltonian for multiband p-d model (\ref{pd_Hamiltonian}) have been performed earlier in paper~\cite{Gav2000} where problem of nonorthogonality of the molecular orbitals of neighboring CuO$_6$ cluster is solved explicitly via diagonalization in k-space~\cite{Raimondi}. To simplify the consideration we assume electron-phonon coupling (\ref{EPI_Hamiltonian}) does not change the set of eigenstates ${\left| p \right\rangle }$ and their energies $E_p$ significantly~\cite{Makarov}. Since the total number of eigenstates is about 100 we proceed with limited set of them that is reasonably if we are interested in low-energy physics. The relevant states are given below. In one-hole sector of Hilbert space formed by d$^9$p$^6$ or d$^{10}$p$^5$ orbital configurations the lowest eigenstates is the spin doublet $\left| {\sigma} \right\rangle $ with ${b_{1g}}$ orbital symmetry. The lowest two-hole states formed by d$^9$p$^5$, d$^{10}$p$^4$ or d$^8$p$^6$ orbital configurations are the $^1{A_1}$ singlet $\left| S \right\rangle $ and $^3{B_{1g}}$ triplet $\left| {{T_m}} \right\rangle $, $m=+1,0,-1$. Therefore matrix elements $\left\langle p \right|c_{f,\mu ,\sigma }^{cell}\left| {p'} \right\rangle $ are defined in the minimal realistic basis $\left\{ {\left| \sigma  \right\rangle ,\left| S \right\rangle ,\left| {{T_m}} \right\rangle } \right\}$. New cell orbitals diagonalizing the unit cell Hamiltonian are the linear combination of initial molecular orbitals~\cite{Gav2000}, the inverse transformation are given by
\begin {equation}
\label{molecular-cell}
c_{\bf{k},\lambda ,\sigma } = \sum\limits_\mu  {{\alpha _{\lambda ,\mu }}{c_{\bf{k},\mu ,\sigma }^{cell}}},
\end {equation}
where index $\mu$ runs trough unmodified $d_{x^2-y^2}$ and $d_{3z^2-r^2}$ orbitals on copper, symmetric $a$ and $b$ atomic orbitals on the plane-oxygen sites and $p_z$ bindng and antibinding orbitals on the apical oxygen.

Taking into account probability coefficients $\left\langle p \right|c_{f,\mu ,\sigma }^{cell}\left| {p'} \right\rangle $ and definition (\ref{molecular-cell}) the electron-phonon interaction (\ref{EPI_Hamiltonian}) is being transformed
\begin{equation}
\label{El_HubbFerm}
{H_{EPI}} = \sum\limits_{{\bf{k}},{\bf{q}},\sigma ,\nu } {\sum\limits_{m,n} {g_{mn}^{SEC}\left( {{\bf{k}},{\bf{q}}{\rm{;}}\nu } \right)\mathop {X_{{\bf{k}} - {\bf{q}}}^m}\limits^ +  X_{\bf{k}}^n\varphi _{\bf{q}}^\nu } }
\end{equation}
with
\begin{equation}
\label{Coeff_El-HubbFerm}
g_{mn}^{SEC} = \sum\limits_{\lambda ,\mu ,\mu '} {{g_\lambda }\left( {{\bf{k}},{\bf{q}}{\rm{;}}\nu } \right)\alpha _{\lambda \mu }^*{\alpha _{\lambda \mu '}}\gamma _{\mu \sigma }^*\left( m \right)} {\gamma _{\mu '\sigma }}\left( n \right).
\end{equation}
Index $m \leftrightarrow \left( {p,p'} \right)$ enumerates
quasiparticle band with intracell energy
$\omega _m = \varepsilon _p \left( {N + 1} \right) - \varepsilon _{p'} \left( N \right)$,
where $\varepsilon_p$ is the $p$-th energy level of the $N$-electron system, this energy acquires the band dispersion due to intercell hoppings~\cite{Gav2000}. It should be stressed that the GTB bands are not free electron bands of the conventional band structure, these bands are formed by quasiparticle excitations between different multielectron terms and the number of states in each particular band depends on the occupation number of the initial and final multielectron configurations, and thus on the electron occupation.

Parameters ${g_{mn}^{SEC}}\left( {{\bf{k}},{\bf{q}}{\rm{;}}\nu } \right)$ of interaction between Hubbard fermions and $\Gamma$-point phonons for $\bf{k}=0$ are presented in Table~\ref{HubbFermPhon}. Analysis of the data in Table~\ref{HubbFermPhon} shows that for each given quasiparticle band index $m$ the strongest interactions are exhibited by A$_{2u}$ modes with 236 and 131~cm$^{-1}$ while E$_g$ modes and E$_u$ ones with 209 and 80~cm$^{-1}$ demonstrate the smallest coupling. It completely consists with the features of electron-phonon interaction determining for a given band index $\lambda$ in Table~\ref{epimat}. However comparison of maximum values of interactions in Tables ~\ref{epimat} and \ref{HubbFermPhon} for each given mode shows their reduction in the limit of strong electron correlation (Table \ref{HubbFermPhon}) approximately on 20-30 percent with insignificant exceptions for two E$_u$ modes. This effect is caused by strong electron correlations in the system that is obviously from equations \ref{El_HubbFerm}, \ref{Coeff_El-HubbFerm},  and \ref{def_Hubb_op}. Moderate suppression of electron-phonon coupling by Coulomb interaction have also been obtained for Holstein-Hubbard model using dynamical mean-field approximation~\cite{Sangiovanni2006}.

Finally we apply density functional theory and generalized tight-binding combined method for electron-phonon coupling study in strongly correlated LSCO-system. The suppression of parameters of interaction between Hubbard fermions and phonons  due to strong electron correlation is demonstrated.


\begin{table*}[htb]
\begin{center}
\begin{tabular}{cc@{\hspace{3mm}}*{3}{c@{\hspace{3mm}}}}
\hline
\hline
       Mode & $\omega$        & \multicolumn{3}{c}{Quasiparticle exitations}\\
            & (cm$^{-1}$)   & $\left( {S, - \sigma ; - \sigma ,S} \right)$
                                         & $\left( {{T_{ \downarrow \left(  \uparrow  \right)}}, - \sigma \left( \sigma  \right); - \sigma \left( \sigma  \right),{T_{ \downarrow \left(  \uparrow  \right)}}} \right)$
                                                      & $\left( {{T_0}, \pm \sigma ; \pm \sigma ,{T_0}} \right)$
                                                                           \\

\hline
 A$_{1g}$ & 415   &  0.688   &  1.492   &   0.746  \\
          & 232   &  0.552   &  0.969   &   0.485  \\[2mm]

A$_{2u}$  & 476   &  1.065   &  1.695   &   0.848  \\
          & 236   &  7.325   &  1.683   &   0.841  \\
          & 131   &  2.661   &  1.899   &   0.950  \\[2mm]

B$_{2u}$  & 227   &  0.035   &  0.403   &   0.201  \\[2mm]

E$_g$     & 209   &  0.085   &  0.111   &   0.056  \\
          &  73   &  0.249   &  0.076   &   0.038  \\[2mm]

E${_u}$   & 727   &  1.158   &  0.503   &   0.252  \\
          & 341   &  1.492   &  0.824   &   0.412  \\
          & 209   &  0.158   &  0.118   &   0.059  \\
          &  80   &  0.145   &  0.177   &   0.089  \\
\hline
\hline
\end{tabular}
\caption{Absolute value of parameters (in eV) of Hubbard fermion-phonon interactions.}
\label{HubbFermPhon}
\end{center}
\end{table*}


\vspace{3mm} \noindent{\bf Acknowledgment.}
This work is partly supported by RFBR (Grants 11-02-00147 and 13-02-01395), Siberian Federal University (Theme F-11), Governmental support of leading scientific schools of Russia (NSh-1044.2012.2), SB-UB RAS project 44, Presidium of RAS program 20.16, Programs of fundamental research of the RAS "Quantum mesoscopic and disordered structures" (12-$\Pi$-2-1002), SB-UB RAS grant 12-C-2-1004 (IAN) and the Dynasty Foundation and ICFPM (EIS).



Figure~1. Comparison of LDA band dispersions for La$_2$CuO$_4$ obtained
within LMTO (dashed lines) and FP-APW+lo (solid lines) methods.
The  Fermi energy corresponds to zero.

Figure~2. The atomic displacements for the three of the A$_{2u}$ modes, which are the vibrations exhibiting the strongest electron-phonon coupling.

\end {document}